\documentstyle[aps,prl,epsf,twocolumn]{revtex}
\begin{document}
\draft
\title{Color-flavor locked strangelets}
\author{Jes Madsen}
\address{Institute of Physics and Astronomy, University of Aarhus, 
DK-8000 \AA rhus C, Denmark}
\date{August 3, 2001; Published in Phys.Rev.Lett.\ 87 (2001) 172003}
\maketitle

\begin{abstract}
Finite lumps of color-flavor locked strange quark matter
(CFL-strangelets) are significantly more stable than strangelets without
color-flavor locking for wide ranges of parameters, increasing the
likelihood of strangelet metastability,
or even absolute
stability beyond some minimum baryon number $A_{\rm min}$.
Whereas bulk CFL strange quark matter is electrically
neutral, CFL-strangelets are positively charged, with 
$Z\approx 0.3 A^{2/3}$. This is quite different from ``ordinary''
strangelets and may provide
a possible test of color-flavor locking if strangelets are detected
in upcoming cosmic-ray space experiments.
\end{abstract}


It has recently been demonstrated that quark matter at high density
may be in a so-called color-flavor locked phase where quarks with different color
and flavor quantum numbers form Cooper pairs with very large binding
energy \cite{alford01}. 
Such a state is significantly more bound than ordinary quark matter, and this
increases the likelihood that quark matter composed of up, down, and strange
quarks may be metastable or even absolutely stable. In other words color-flavor
locked quark matter rather than nuclear matter may be the ground state of
hadronic matter. The present Letter discusses this possibility with particular
emphasis on the consequences for finite size quark matter lumps, so-called
strangelets.

As shown by Rajagopal and Wilczek\cite{rajwil2001}
the color-flavor locked phase is electrically neutral in bulk for a significant
range of chemical potentials and s-quark mass. The reason for the
neutrality is that
BCS-like pairing minimises the energy if the quark Fermi-momenta are equal
(since pairing happens between quarks of different color and flavor, and 
opposite momenta $\vec p$ and $-\vec p$). For equal Fermi-momenta, the number of
up, down and strange quarks are equal, and the net charge of the
system is zero without any need for (or even room for) electrons.
This has important consequences for the physics of compact stars
containing color-flavor locked quark matter\cite{alfal2001}.
As demonstrated below, it also has important consequences for the mass
and charge properties of finite lumps of strange quark matter
(strangelets) if these consist of color-flavor locked rather than
``normal'' quark matter.

For ``normal'' quark matter it is known that finite size effects
increase the energy
\cite{bodmer71,chin79,witten84,farhi84,berger87,madsen93a,gilson93,madsen93b,madsen94,schaff97,madsen2000}, 
and make the charge more positive
\cite{madsen2000}. As shown below a similar thing happens in
color-flavor locked quark matter. Even for quark matter where equal
Fermi momenta are enforced by color-flavor locking, the finite-size
effects change the net quark charge of the system from zero to a
positive value. The resulting charge--mass relation ($Z\approx
0.3A^{2/3}$) differs significantly from ordinary strangelets and may
allow for an experimental test of color-flavor locking.

Following Refs.\ \cite{rajwil2001,alfal2001}, the color-flavor locked
phase with full pairing has a free energy (volume terms
only)
\begin{equation}
\Omega_{{\rm CFL},V}V=\Omega_{{\rm
free},V}(\mu_u,\mu_d,\mu_s,p_{Fu},p_{Fd},p_{Fs})V + \Omega_{{\rm pair},V}V
\end{equation}
where $\Omega_{{\rm pair},V}\approx -3\Delta^2\mu^2/\pi^2$, with
$\Delta\approx$10--100 MeV being the pairing energy gap and $\mu$ the
average quark chemical potential.
$\Omega_{{\rm free},V}$ is the usual Fermi-gas result (see below) as a
function of the individual quark chemical potentials, $\mu_i$, and
Fermi-momenta, $p_{Fi}$, but for
the color-flavor locked phase the Fermi-momenta are all equal,
$p_{Fu}=p_{Fd}=p_{Fs}\equiv p_F$ \cite{rajwil2001}. 

Finite lumps of quark matter have additional contributions to
$\Omega_{\rm free}$ in terms of surface and curvature energies. 
Presumably, also $\Omega_{\rm pair}$ would contain finite-size terms,
but these are likely to be small compared to the corrections to 
$\Omega_{\rm free}$, as long as $\Omega_{\rm pair}$ itself is a
perturbation to $\Omega_{\rm free}$.
Here such higher order terms will be neglected, and it will be assumed
that the CFL state keeps all quark Fermi
momenta equal to optimize the pairing energy. The common value of $p_F$
is found by minimizing $\Omega_{\rm free}$ at fixed radius, and the
equilibrium radius is found by minimizing the total energy
with respect to radius at fixed baryon number.

The calculation of the unpaired contributions, $E$, to the total energy,
$E_{\rm CFL}= E+\Omega_{\rm pair}$,
is performed within the MIT bag model\cite{degrand75}, with $\alpha_S=0$.
Here the energy of a system composed of
quark flavors $i$ is given by
$E=\sum_i(\Omega_i+N_i\mu_i)+BV$,
where $\Omega_i$, $N_i$ and $\mu_i$ 
denote thermodynamic potentials,
total number of quarks, and chemical potentials, respectively. $B$ is
the bag constant, $V$ is the bag volume.
The thermodynamical quantities can be derived from a density of states of
the form \cite{balian70}
$
{{dN_i}\over{dk}}=6 \left\{ {{k^2V}\over{2\pi^2}}+f_S\left({m_i\over
k}\right)kS+f_C\left({m_i\over k}\right)C+ .... \right\} ,
$
where a sphere has area $S=4\pi R^2$ and curvature $C=8\pi R$.
The functions $f_S$ and $f_C$ depend on the boundary conditions.
For the MIT-bag model
$f_S(m/k)=-\left[ 1-(2/\pi)\tan^{-1}(k/m)\right] /8\pi$\cite{berger87}
and $f_C(m/k)=
\left[ 1-3k/(2m)\left(\pi/2-\tan^{-1}(k/m)\right)\right]/12\pi^2$
\cite{madsen94}.
The number of quarks of flavor $i$ is
$
N_i=\int_0^{p_{Fi}}({dN_i}/{dk})dk=n_{i,V}V+n_{i,S}S+n_{i,C}C,
$
and the corresponding thermodynamic potentials are
$\Omega_i=\int_0^{p_{Fi}}({dN_i}/{dk})(\epsilon_i(k)-\mu_i)dk=
\Omega_{i,V}V+\Omega_{i,S}S+\Omega_{i,C}C,$ 
where $\epsilon_i(k)=(k^2+m_i^2)^{1/2}$.
The expressions obey $\partial\Omega_i/\partial\mu_i=-N_i$, and
$\partial\Omega_{i,j}/\partial\mu_i=-n_{i,j}$. 

With $\lambda_i\equiv m_i/p_{Fi}$ this gives
\begin{eqnarray}
\Omega&&_{i,V}=-{3{p_{Fi}^4}\over {8\pi^2}}\left( 
{{8\mu_i}\over{3 p_{Fi}}}-2(1+\lambda_i^2)^{3/2}
+\lambda_i^2(1+\lambda_i^2)^{1/2}\right.\cr
&&\left. +\lambda_i^4\ln{{1+(1+\lambda_i^2)^{1/2}}\over\lambda_i}
\right) ,
\end{eqnarray}
and
\begin{equation}
n_{i,V}=p_{Fi}^3/\pi^2 .
\end{equation}

Massless quarks have $\Omega_{i,S}=0$, with a significantly more
cumbersome equation for s-quarks.
The corresponding change in quark number per unit area,
\begin{eqnarray}
n_{i,S}&&=-{3\over{4\pi}}p_{Fi}^2\left[{1\over 2}+{{\lambda_i}\over{\pi}}
-{1\over{\pi}}\left( 1+\lambda_i^2\right)\tan^{-1}(\lambda_i^{-1}) \right] ,
\end{eqnarray}
is always negative, approaching 0
for $\lambda_i\rightarrow 0$ (massless quarks) and $-3p_{Fi}^2/8\pi$ for
$\lambda_i\rightarrow \infty$.

For massless u and d quarks $\Omega_{i,C}=(2p_{Fi}\mu_i-p_{Fi}^2)/
(8\pi^2)$, and $n_{i,C}=-p_{Fi}/(4\pi^2)$. For a massive quark
\begin{eqnarray}
n_{i,C}={{p_{Fi}}\over{8\pi^2}}\left[ 1-{{3\pi}\over{2\lambda_i}}
+3{{1+\lambda_i^2}\over{\lambda_i}}\tan^{-1}(\lambda_i^{-1})\right] .
\end{eqnarray}

Quark matter in weak equilibrium has $\mu_s=\mu_d=\mu_u+\mu_e$,
maintained by reactions like $u+d\leftrightarrow s+u$, $u+e^-
\leftrightarrow d+\nu_e$.
For a system with zero electron chemical potential, e.g.\ a finite strangelet
with $A\ll 10^7$ (for $A>10^7$ the strangelet radius exceeds the
electron Compton wavelength, so that some electrons may be trapped inside
the quark phase),
weak interactions keep an equilibrium characterised by $\mu_u=\mu_d=\mu_s
\equiv\mu$. 

The surface and curvature terms have two important implications for 
the color-flavor locked strangelets.
First, as for normal strangelets, they destabilize small systems
relative to bulk quark matter.
Second and perhaps more important, quark lumps in the color-flavor
locked phase with equal Fermi-momenta are no longer electrically
neutral as in bulk, but instead gain a net positive quark charge due
to the relative suppression of s-quarks for fixed $p_{F}$,
similar to the charge increase which has been pointed out 
for normal quark matter\cite{madsen2000}.

In bulk, equal Fermi-momenta means
equal quark numbers since the pairing terms are
equal\cite{rajwil2001}, $N_{i,{\rm pair}}=V2\Delta^2\mu /\pi^2$,
and the contribution from $\Omega_{\rm free}$
is $N_{i, {\rm free}}= V p_{F}^3/\pi^2$. But for finite systems, the
number of massive quarks is generally reduced relative to massless
quarks at fixed $p_{F}$. This is explicitly the case for the MIT bag
model boundary conditions, where surface tension is related to a
suppression of the s-quark wave-function at the surface. MIT bag
boundary conditions correspond to no flux of quarks across the
surface. For a massive s-quark this lowers the s-quark density near
the surface relative to that of u and d ($n_{s,S}<0$), 
leading to a net increase in
the total electrical charge, which therefore must become positive for
quark matter with equal Fermi-momenta, i.e.\ color-flavor locked quark
matter. 
This is a consequence of quantum
mechanics with validity beyond the MIT bag model. 
In the nonrelativistic limit of a very massive quark 
the wave function must be zero at the boundary. This is not the case
for a relativistic quark. Therefore one would generally expect massive
s quarks to be more suppressed than u and d.

\begin{figure}
\epsfxsize=8.5cm\epsfbox{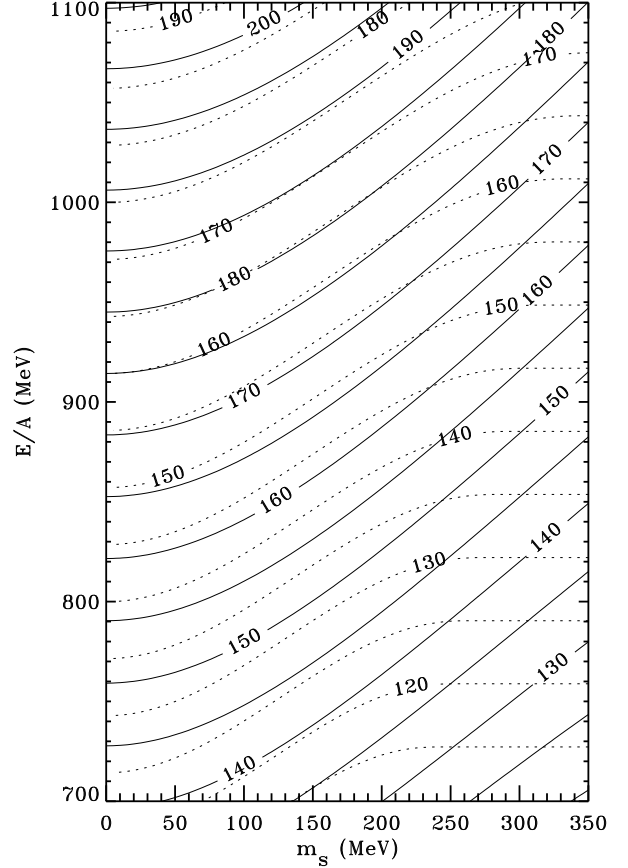}
\caption{
Energy per baryon in MeV as a function of $m_s$
for bulk strange quark matter with $B^{1/4}$ in MeV as indicated. Full lines are
results for color-flavor locked quark matter ($\Delta=100$MeV), dashed
lines without color-flavor locking.
}
\label{fig1}
\end{figure}

Figure 1 shows energy per baryon for bulk CFL-quark matter 
with $\Delta=100$~MeV as a function of bag constant and s-quark mass.
Results for non-CFL quark matter are shown for comparison. As
demonstrated previously (e.g.\ \cite{alford01,rajwil2001,alfal2001}) 
it is evident
that color-flavor locking significantly lowers the energy and makes
quark matter more stable. Up to a hundred MeV per baryon is gained for
some parameter choices. This means that strangelets, if
they are color-flavor locked, may be absolutely stable 
($E/A<930$~MeV) for bag constants
$B^{1/4}$ as high as 180~MeV for low strange quark mass, 170~MeV
for $m_s\approx 150$~MeV, and 160~MeV for $m_s\approx 300$~MeV, whereas
stability without color-flavor locking requires $B^{1/4} < 163$~MeV even
for massless s-quarks. The corresponding ranges for metastability
are pushed to higher $B$ as well. For normal quark matter
$B^{1/4}$ is bounded from below by the value 146~MeV, below which
ordinary nuclei would be unstable against spontaneous decay into
two-flavor up-down quark matter. For $\Delta=100$~MeV this bound
increases to $B^{1/4}>156$~MeV to avoid spontaneous nuclear decay into a
two-flavor color superconducting state where two of the three colors of
up and down quarks form Cooper pairs \cite{note23}. 
The proper value of $B$ is an important but unsettled issue. The
original MIT bag model fits to hadrons implied $B^{1/4}=145$~MeV,
but this depends significantly on the choice of $m_s$, $\alpha_s$, and
not least a phenomenological zero-point energy, which is negligible for
larger strangelets. On the other hand, a naive comparison with lattice
determinations of the quark-hadron phase transition temperature at
$\mu=0$ (circumstances very different from what the bag model was
created to describe) would argue for a higher value. Using the bag model result
$B^{1/4}\approx 1.45 T_c$ and the lattice results for $T_c$ one finds that
$B^{1/4}>200$~MeV. In the latter case, a value of $\Delta$ in excess of the
100~MeV considered here would be necessary for CFL-strangelet stability.

\begin{figure}
\epsfxsize=8.5cm\epsfbox{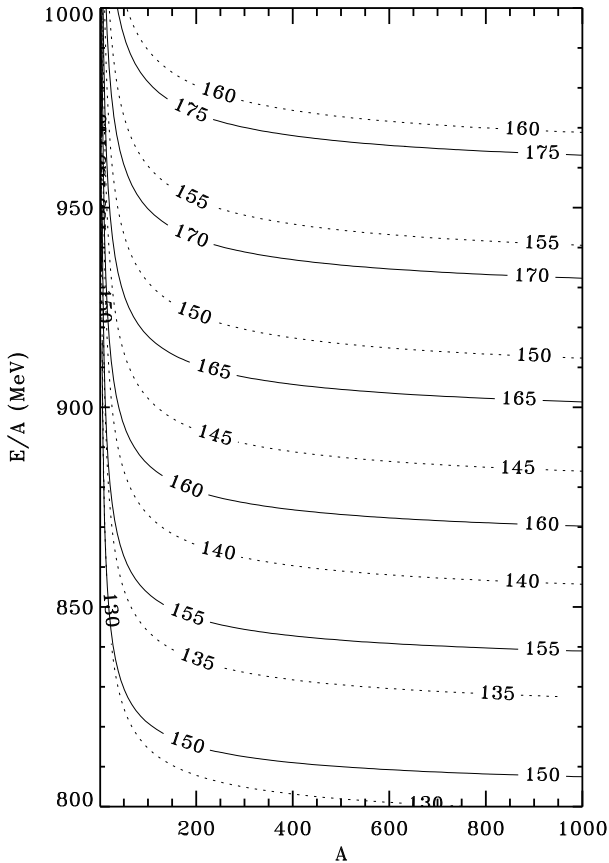}
\caption{
Energy per baryon in MeV as a
function of $A$ for CFL-strangelets (full curves) and ordinary
strangelets (dashed curves) with $B^{1/4}$ in MeV as indicated,
$\Delta=100$MeV and $m_s=150$MeV.
}
\label{fig2}
\end{figure}

Figure 2 shows energy per baryon for finite size strangelets with and
without color-flavor locking.
For a given $B$ the strangelet mass-per-baryon 
increases dramatically for low $A$
because of the surface and curvature energies, and there is a lower
bound $A_{\rm min}$ for the baryon number of stable (metastable),
strangelets, which depends on parameters. The overall gain in stability
for CFL-strangelets relative to non-CFL strangelets mainly reflects the
gain in bulk binding.

\begin{figure}
\epsfxsize=8.5cm\epsfbox{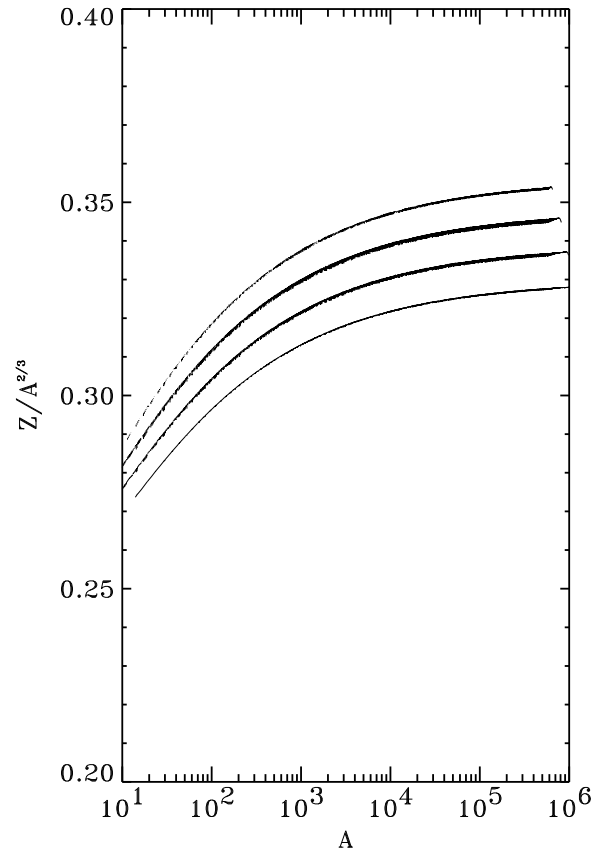}
\caption{
Charge divided by $A^{2/3}$ as a
function of $A$ for CFL-strangelets with $B^{1/4}=150$, 160, 170, and
180 MeV (top to bottom), $\Delta=100$MeV and $m_s=150$MeV.
}
\label{fig3}
\end{figure}

Perhaps more interesting, and in significant contrast to the properties
of non-CFL strangelets, is the charge $Z$ of strangelets as a function of
$A$. Since bulk CFL strange quark matter is electrically neutral
\cite{rajwil2001,alfal2001}, the total charge in the case of finite
strangelets with $\mu_e=0$ comes from the surface suppression of massive
relative to massless quarks following from the MIT boundary conditions
(and more generally from the fact that $f_S$ is negative for a massive
quark). As can be seen from Fig.\ 3, the charge per baryon is almost
independent of $B$ and it is only weakly dependent on $\Delta$ and $m_s$. The
characteristic value is $Z\approx 0.3 A^{2/3}$, which can be understood
as follows. The main charge contribution comes from $n_{s,S}$ as
$Z\approx -(1/3)n_{i,S}S\approx (2/\pi )(m_s/p_F)(p_F R)^2$ (expansion
for small $m_s$), saturating at $0.5 (p_F R)^2$ for high $m_s$.
Neglecting the pairing contribution to $A$, $A\approx Vp_F^3/\pi^2=
4(p_F R)^3/3\pi$, so the $Z\propto A^{2/3}$ behavior is clear, and
the prefactor is not very parameter dependent.
For non-CFL strangelets, the charge is
the volume charge density times volume for small $A$, and therefore
proportional to $A$ itself ($Z\approx 0.1 A$), 
until the system becomes larger than the
Debye screening length ($\approx 5$~fm; $A\approx 150$), 
beyond which the charge is
mainly distributed within a Debye length from the surface,
and $Z\approx 8A^{1/3}$\cite{farhi84,berger87,heiselberg,note}. 
Charge screening is negligible
for CFL-strangelets because the net charge is already concentrated near
the surface, as it is enforced by the surface boundary conditions
(in spite of the fact, that the quarks share a common Fermi momentum),
whereas the central regions remain charge neutral.

The unusual charge properties, in particular the small 
$Z/A$-ratio, has long been recognized as a crucial signature
for experimental identification of strangelets. The results above show
that the actual ($A$,$Z$) relation is very different in the case of
CFL-strangelets ($Z/A\propto A^{-1/3}$) as compared to ordinary
strangelets ($Z/A$ constant for small $A$ and $\propto A^{-2/3}$ for
large $A$, reaching $Z/A\propto A^{-1/3}$ only asymptotically). 
This is an important distinction which calls for a
re-analysis of some of the limits derived in previous experimental
strangelet searches\cite{exp}. 
If strangelets are detected in upcoming cosmic ray
space experiments (in particular, the Alpha Magnetic Spectrometer\cite{ams} 
and ECCO\cite{ecco} will be sensitive to strangelets in different charge
(mass) regimes), 
the difference may also allow an
experimental test of color-flavor locking in quark matter.

This work was
supported in part by the Theoretical Astrophysics Center under the
Danish National Research Foundation, and by the Danish Natural Science
Research Council.


\begin{references}
\bibitem{note23}
Notice that the introduction of
pairing does not just act like a rescaling of $B$ with equal consequences
for 2- and 3-flavor systems. This is because pairing in a two flavor
system involves only 4/9 of the available degrees of freedom in a three
flavor system. Thus, the range of allowed $B$-values corresponding to
CFL-strangelet stability or metastability is wider than for ordinary strange
quark matter.
\end{references}
\end{document}